\documentstyle[preprint,tighten,aps,epsfig]{revtex}
\begin{document}

\title{$nd$ scattering lengths from a quark-model based $NN$ interaction}

\author{H. Garcilazo$^{(1)}$ and A. Valcarce$^{(2)}$}
\address{$(1)$ Escuela Superior de F\'\i sica y Matem\'aticas, 
Instituto Polit\'ecnico Nacional, 
Edificio 9, 07738 M\'exico D.F., Mexico}
\address{$(2)$ Departamento de F\'\i sica Fundamental, 
Universidad de Salamanca, E-37008 Salamanca, Spain}
\maketitle

\date{\today}
\begin{abstract}
We calculate the doublet and quartet neutron-deuteron scattering lengths
using a nonlocal nucleon-nucleon interaction fully derived from quark-quark
interactions. We use as input the $NN$ $^1S_0$ and 
$^3S_1$-${}^3 \! D_1$ partial waves. Our result for the 
quartet scattering length agrees well with the experimental value while
the result for the doublet scattering length does not. However, 
if we take the result for the doublet scattering length together with the 
one for the triton binding energy they agree well 
with the so-called Phillips line.
\end{abstract}

\maketitle

In a previous paper~\cite{Jul02} the properties of the 
bound state of three nucleons, the triton,
were studied using a nonlocal nucleon-nucleon ($NN$) interaction
obtained from the chiral quark model~\cite{Val05}.
In this Brief Report we would like to complete that study
by calculating within the same framework the neutron-deuteron
doublet and quartet scattering lengths.

The nonlocal nucleon-nucleon interaction, obtained
from the chiral quark model, is based on
a Lippmann-Schwinger formulation of the resonating group method in
momentum space. A detailed description of the 
method and a summary of the parameters of the model is given in
Ref.~\cite{Jul02}. In this reference it was shown that this
$NN$ interaction describes very
well the properties of the deuteron as well as the $^1S_0$ and
$^3S_1$-${}^3\!D_1$ phase shifts.
Besides, it was found that the nonlocal $NN$ interaction
obtained from the chiral quark model predicts a triton binding
energy of 7.72 MeV, comparable to the predictions by 
conventional meson-exchange models. Thus, it is important to see 
if that result extends also to the neutron-deuteron scattering
lengths.

We will write down the integral equations that determine the $nd$
scattering lengths using the partial-wave basis states

\begin{equation}
\mid p_iq_i;\rho_i\rangle\equiv\mid p_iq_i;\ell_is_ij_ii_i\lambda_iJ_i\rangle,
\end{equation}
where if $\sigma _i$ and $\tau _i$ stand for
the spin and isospin of particle $i$ then
$\ell_i$, $s_i$,
$j_i$, $i_i$, $\lambda_i$, and $J_i$ are the orbital angular momentum,
spin, total angular momentum, and isospin of the pair $jk$ while
$\lambda_i$ is the orbital angular momentum between particle
$i$ and the pair $jk$ and $J_i$ is the result of coupling $\lambda_i$
and $\sigma_i$. The conserved quantum numbers are $J$,
the total angular momentum and $I$, the total isospin.

The Faddeev equations that determine the $nd$ scattering lengths in
the special case when one considers only $S$ wave configurations were
written down in Ref.~\cite{Gar85}. The generalization of these 
equations to the case when one includes arbitrary orbital angular 
momenta are

\begin{eqnarray}
\langle p_iq_i;\rho_i\mid T_i^{JI}\mid\phi_0\rangle & = & 
2\delta_{s_i 1}\delta_{j_i 1}\delta_{i_i 0}\delta_{\lambda_i 0}
{1\over q_i^2}\delta(q_i)
G_0^{-1}(E;p_iq_i)\phi_{\ell_i}(p_i) 
\nonumber \\
& + & \sum_{\ell_i^\prime\rho_j}
\int_0^\infty q_j^2 dq_j\int_{-1}^1 d{\rm cos}\theta \,
t_{\ell_i\ell_i^\prime s_i j_i i_i}(p_i,p_i^\prime;E-{3q_i^2
\over 4M})
\nonumber \\
& \times & G_0(E;p_jq_j)
D_{ij;JI}^{\rho_i^\prime\rho_j}(q_i,q_j,{\rm cos}\theta)
\langle p_jq_j;\rho_j\mid T_j^{JI}\mid\phi_0\rangle , 
\label{for20}
\end{eqnarray}
where $t_{\ell_i\ell_i^\prime s_i j_i i_i}(p_i,p_i^\prime;e)$
are the nucleon-nucleon $t$-matrices,
$M$ is the mass of the nucleon, and $E=-B_d$ where  $B_d$ is the 
binding energy of the deuteron. 
$\phi_{\ell_i}(p_i)$ is the deuteron wave function with orbital 
angular momentum $\ell_i$, and

\begin{equation}
G_0(E;p_iq_i)={1\over E -p_i^2/M-3q_i^2/4M + i\epsilon},
\label{for21}
\end{equation}

\begin{equation}
p_i^\prime=\sqrt{q_j^2+q_i^2/4+q_iq_j{\rm cos}\theta},
\label{for22}
\end{equation}

\begin{equation}
p_j=\sqrt{q_i^2+q_j^2/4+q_iq_j{\rm cos}\theta},
\label{for23}
\end{equation}

\begin{equation}
\rho_i^\prime\equiv\{\ell_i^\prime s_ij_ii_i\lambda_iJ_i\}.
\label{for24}
\end{equation}
$D_{ij;JI}^{\rho_i^\prime\rho_j}(q_i,q_j,{\rm cos}\theta)$ are the
angular momentum-spin-isospin recoupling coefficients which are
defined by Eqs. (21)-(26) of Ref. \cite{Gar07}.

After obtaining the solution of Eqs. (\ref{for20}) the neutron-deuteron
scattering lengths are calculated as

\begin{eqnarray}
a_J & = & {2M\pi\over 3}\sum_{\rho_j}\sum_{\rho_i}
\delta_{s_j 1}\delta_{j_j 1}\delta_{i_j 0}\delta_{\lambda_j 0}
\int_0^\infty q_i^2 dq_i
\nonumber \\
& \times & \phi_{\ell_j}(q_i) 
D_{ji;J{1\over 2}}^{\rho_j\rho_i}(0,q_i,0)
\langle q_i/2,q_i;\rho_i\mid T_i^{J{1\over 2}}\mid\phi_0\rangle .
\label{for25}
\end{eqnarray}

Since in Ref.~\cite{Jul02} the triton binding energy was calculated
using as input the $NN$ $^1S_0$ and 
$^3S_1$-${}^3 \! D_1$ partial waves we will use the same prescription in
our calculation of the neutron-deuteron scattering lengths.
This implies a five-channel Faddeev
calculation for the doublet scattering length $a_{1/2}$
and a seven-channel 
calculation for the quartet scattering length $a_{3/2}$.
We give these channels in Table \ref{t1}.

Our method~\cite{Gar07} to solve the Faddeev equations 
consists in transforming them from being integral 
equations in two continuous variables into integral equations in just one
continuous variable. This is achieved by 
expanding the two-body $t-$matrices in terms of Legendre polynomials as
\begin{equation}
t_{i}(p_i,p^\prime_i;e)=\sum_{nr}P_n(x_i)
\tau_{i}^{nr}(e)P_r(x^\prime_i),
\label{for11}
\end{equation}
where $P_n$ and $P_r$ are Legendre polynomials,

\begin{equation}
x_{i}={\frac{p_{i}-b}{p_{i}+b}},  \label{for9}
\end{equation}

\begin{equation}
x_{i}^\prime={\frac{p_{i}^\prime-b}{p_{i}^\prime+b}},  \label{for9p}
\end{equation}
and $p_i$ and $p_i^\prime$ are the initial and final relative momenta of
the pair $jk$ while $b$ is a scale parameter on which the results do
not depend. We found that using
$b=3$ fm$^{-1}$ leads to very stable results while for the expansion
(\ref{for11}) we found convergence with
twelve Legendre polynomials, i.e., $0\le n\le 11$.

We have calculated also the triton binding energy to 
check that our calculation reproduces the value 
obtained in Ref.~\cite{Jul02}. We give in Table \ref{t2}
our results for the triton binding energy and the two
neutron-deuteron scattering lengths as well as the 
corresponding experimental results~\cite{Dil71}.  
As pointed out in Ref.~\cite{Jul02}, the result $B=7.72$ MeV
for the triton binding energy predicted by the chiral quark model 
is comparable to the values obtained by conventional meson-exchange
models like Nijmegen or Bonn~\cite{Sch00}
since the theoretical value differs by
less than 1 MeV from the experimental result. The situation in the 
case of the doublet scattering length $a_{1/2}$ appears to be
somewhat worse since the theoretical value 1.13 fm is almost a factor
of two larger than the experimental result. However, as we will see
next, that is not the case. As it is well-known~\cite{Phi68}, 
the results obtained from a given theoretical model for the triton
binding energy $B$ and the doublet scattering length $a_{1/2}$
are strongly correlated. The results obtained from different models
follow what is known as the Phillips
line which is a straight line relating $B$ versus $a_{1/2}$. 
For example, in Ref.~\cite{Ben81} a five-channel calculation
similar to ours was performed using three local 
potentials~\cite{Mal69,Det73,Rei68}; therefore, we made 
a minimum-square fit of their results for $B$ and $a_{1/2}$
to obtain the Phillips line $a_{1/2}=6.352-0.677B$ which
for $B=7.72$ MeV gives $a_{1/2}=1.13$ fm in agreement with
our result. The most complete calculations of 
$B$ and $a_{1/2}$ have been performed in Ref.~\cite{Wit03} 
for a variety of modern nucleon-nucleon force models 
\cite{Wir95,Mac01,Sto94} and
where they have included the higher angular momentum two-body channels 
as well as three-body forces \cite{Pud97,Fri99,Coo01,Coo79}.
Using the values of
$B$ and $a_{1/2}$ for the 48 different models considered in
Ref. \cite{Wit03} we obtained the Phillips line 
$a_{1/2}=7.028-0.756B$ which for
$B=7.72$ MeV will give a doublet scattering length 
$a_{1/2}=1.19$ fm which is also quite close to the result of our 
calculation. Our result for the quartet scattering length 
$a_{3/2}=6.40$ fm  agrees well with experiment 
and with Refs.~\cite{Ben81,Wit03}. 

The reason why our calculation agrees well with experiment in
the case of the $J=3/2$ channel but not in the case of the $J=1/2$
channel is that the former is determined by the pure $S$-wave 
configuration $\ell=\lambda=0$ while the latter has important contributions 
from higher angular momentum two-body channels and from three-body
forces, both of which are lacking into our model. This is very
similar to the situation encountered in effective field theory where
the $J=3/2$ channel is very well explained by $S$-wave models without
three-body forces~\cite{Bed98,Gri05} 
while for the $J=1/2$ channel the theory can 
produce sensible results only when a three-body force is 
included~\cite{Bed00,Pla06}.

In summary, 
we conclude that the nonlocal $NN$ interaction obtained from the 
chiral quark model gives results for the neutron-deuteron
scattering lengths which are comparable to those obtained from
conventional meson-exchange models.

\vspace*{1cm} \noindent
This work has been partially funded by Ministerio de Educaci\'{o}n y Ciencia
under Contract No. FPA2007-65748 and by COFAA-IPN (M\'{e}xico).

\begin{table}[tbp]
\caption{Three-body channels 
that contribute to
a given $NNN$ state with total isospin $I$ and 
total angular momentum $J$.}
\label{t1}
\begin{tabular}{|cccccccc|}
$I$ & $J$ 
& $\ell_i$ & $s_i$ &  $j_i$ &  $i_i$ & $\lambda_i$ &  $J_i$ \\
\tableline 
0 & 1/2 
  & 0 & 0 & 0 & 1/2 & 0 & 1/2 \\
& & 0 & 1 & 1 & 1/2 & 0 & 1/2 \\
& & 2 & 1 & 1 & 1/2 & 0 & 1/2 \\
& & 0 & 1 & 1 & 1/2 & 2 & 3/2 \\ 
& & 2 & 1 & 1 & 1/2 & 2 & 3/2 \\ 
  \\ \hline 
0 & 3/2 
  & 0 & 0 & 0 & 1/2 & 2 & 3/2 \\
& & 0 & 1 & 1 & 1/2 & 0 & 1/2 \\
& & 2 & 1 & 1 & 1/2 & 0 & 1/2 \\
& & 0 & 1 & 1 & 1/2 & 2 & 3/2 \\ 
& & 0 & 1 & 1 & 1/2 & 2 & 5/2 \\ 
& & 2 & 1 & 1 & 1/2 & 2 & 3/2 \\ 
& & 2 & 1 & 1 & 1/2 & 2 & 5/2 \\ 
\end{tabular}
\end{table}

\begin{table}[tbp]
\caption{Triton binding energy $B$ (in MeV) and scattering lengths
$a_{1/2}$ and $a_{3/2}$ (in fm)  as compared with the 
corresponding experimental values.}
\label{t2}
\begin{tabular}{|ccc|}
Quantity &  Theory & Experiment \\
\tableline 
$B$ & 7.72 & 8.48 \\
$a_{1/2}$ & 1.13 & 0.65$\pm$0.04 \\
$a_{3/2}$ & 6.40 & 6.35$\pm$0.02 \\
\end{tabular}
\end{table}

\end{document}